\documentclass[twocolumn,showpacs,preprintnumbers,amsmath,amssymb]{revtex4}
\usepackage{graphicx}
\usepackage{dcolumn}
\usepackage{bm}

\newcommand{\ie}{ i.e.,~}
\newcommand{\eg}{ e.g.,~}
\newcommand{\etal}{{\it et al.~}}

\newcommand{\half}{\frac{1}{2} }

\def\deriv#1{\frac{\partial}{\partial#1}}
\def\der#1#2{\frac{\partial#1}{\partial#2}}

\begin{document}

\title{Understanding spin transport from the motion in SU(2)$\times$U(1) fields}

\author{Pei-Qing Jin and You-Quan Li}
\affiliation{Zhejiang Institute of Modern Physics and Department of Physics,
Zhejiang University, Hangzhou 310027, People's Republic of China }

\begin{abstract}
Starting from a continuum constituted by charged tops,
we formulate the classical counterpart of a previously obtained
covariant continuity-like equation for the spin current.
Such a formulism provides an intuitive picture
to elucidate the non-conservation of the spin current
and to interpret the condition for the emergence of
an infinite spin relaxation time.
It also facilitates the discussion on the spin precession in
a one-dimensional quantum wire with Dresselhaus
and Rashba spin-orbit couplings.
Furthermore, we derive the diffusion equations
for both the charge and spin densities
and find that they couple to each other due to the \emph{Zitterbewegung}.
\end{abstract}

\pacs{72.25.Dc, 72.25.-b, 03.65.-w, 85.75.-d}

\received{January 16, 2008}

\maketitle

\section{Introduction}

Recently, the coherent manipulation of the spin degree of
freedom in semiconductor, one of the main subjects in
spintronics~\cite{Zutic}, has attracted much attention
for its potential applications in future information
precessing and storage technologies.
Understandings on the basic properties of the
spin transport, spin dynamics and spin relaxation
in semiconductors are required for the design of
spin-based devices.
In this field, the intrinsic spin-orbit coupling (SOC)
in semiconductors is considered to be an effective route
for manipulating the spin degree of freedom
and generating the spin current.
The former aspect is involved in the
proposal of Datta-Das spin transistor~\cite{Datta-Das},
while the latter one leads to the numerous
studies of the spin Hall effect both
theoretically~\cite{Daykonov,Hirsch,ZhangSC,Niu0403} and
experimentally~\cite{Wunderlich,Awschalom,Sih,Zhao,Tinkham}.
However, the SOC also brings about
the non-conservation of the spin current, which makes
the definition of the spin current cumbersome~\cite{Li-current,Niu05,Ma}.
A covariant form for the continuity-like
equation for the spin current was given~\cite{Li-current}
in the terminology of SU(2) gauge potentials.
It was shown to play an essential role
in guaranteeing the consistency of a generalized
Kubo formula for the linear response to non-Abelian
fields with different gauge fixings~\cite{Li-Kuboformula}.
As the work mentioned above are all based on
quantum mechanics, one may ask
what is the classical interpretation for
the non-conservation of spin current
and whether this classical picture can bring new clues
to the understanding on the spin transport?

In this paper, we investigate the motion of electrons in the
presence of SOCs (regarded as SU(2) fields) as well as the U(1)
electromagnetic fields.
Considering a continuum constituted by charged tops,
we obtain the classical counterpart of
the continuity-like equation for the spin current.
It takes the same form as that proposed in the view of the quantum
mechanics~\cite{Li-current} and exhibits an intuitive picture for
the non-conservation of the spin current. This classical picture
of the motion of electrons is the basis of our discussions on the
spin transport in present paper.
The whole paper is organized as follows.
In Sec.~\ref{sec:continuity-like equation}, the classical
counterpart of the continuity-like equation for the spin current
is formulated. In
Sec.~\ref{sec:spin precession}, we discuss the precession of the
spin orientation in a ballistic quantum wire with both Dresselhaus
and Rashba SOCs.
In Sec.~\ref{sec:single spin}, we derive the equations of motion
for a single spin, from which we can get the condition for the
emergence of infinite spin relaxation time.
Starting with the semiclassical Boltzmann equation,
we derive the diffusion equations for both the
charge and spin densities in Sec.~\ref{sec:diffusion equation}.
Finally, a brief summary is given in Sec.~\ref{sec:summary}.

\section{Classical counterpart of continuity-like equation}\label{sec:continuity-like equation}

We start with considering a moving top
(classical analogy of spin)
which rotates at a certain rate.
A continuum constituted by such
kind of tops is completely characterized
by a local velocity field $\mathbf{v}(\mathbf{r}, t)$,
a local particle-density field
$\rho(\mathbf{r}, t)$ together with a local
alignment field $\vec{N}(\mathbf{r}, t)$ \cite{footnote}.
Hereafter, a letter in boldface denotes for a vector
in the conventional spatial space with indices $i,j=1,2,3$
labelling its components, \eg $\mathbf r=x_i \hat e_i$
where $\hat e_i$ are the bases and repeated indices
are summed over;
while the one with an arrow over head is a vector
in the spin space, \eg $\vec N=(N^x, N^y, N^z)$.
\begin{figure}[h]
\includegraphics[width=66mm]{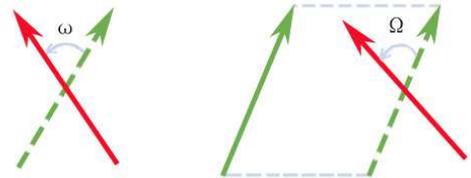}
\caption{\label{fig:procession} (Color on line)
The left scheme depicts the time-evolution
of $\vec{N}$ at a point $\mathbf{r}$;
the right scheme illustrates the spatial deviation
of $\vec{N}$ by comparing the fields
at two neighborhood points $\mathbf{r}$
and $\mathbf{r}+\Delta \mathbf{r}$
for which the parallel displacement is inevitable.}
\end{figure}
The time evolution of $\vec{N}(\mathbf{r}, t)$
is determined by comparing $\vec{N}$
at different times at the same place, \ie
$\vec{N}(\mathbf{r}, t+\Delta t)-\vec{N}(\mathbf{r}, t)
 =\vec{\omega}\times \vec{N}(\mathbf{r}, t)\Delta t$,
while its spatial deviation is determined by comparing
$\vec{N}$ at different places simultaneously, \ie
$\vec{N}(\mathbf{r}+\Delta \mathbf r, t)-\vec{N}(\mathbf{r},t)
 =\vec{\Omega}_i\times \vec{N}(\mathbf{r}, t)\Delta x_i$.
The vector fields $\vec \omega$ and $\vec \Omega_i$
are natural consequences of $\vec{N}$ being a vector
with constant module, then we have
\begin{eqnarray}\label{eq:omegafields}
\deriv{t}\vec{N}(\mathbf{r}, t)
  =\vec{\omega}(\mathbf{r}, t)\times\vec{N}(\mathbf{r}, t),
                   \nonumber\\
\deriv{x_i}\vec{N}(\mathbf{r}, t)
 =\vec{\Omega}_i(\mathbf{r}, t)\times\vec{N}(\mathbf{r}, t).
\end{eqnarray}
By making use of these two relations
together with the density conservation
$\displaystyle \der{\rho}{t}+\der{j^{~}_i}{x_i}=0$,
we can easily derive a continuity-like equation
\begin{equation}\label{eq:continuty-like}
\bigl(\deriv{t}-\vec{\omega}\times\bigr)\vec{\sigma}
 +\bigl(\deriv{x_i}-\vec{\Omega}_i\times\bigr)\vec{J}_i=0,
\end{equation}
as long as the natural definitions of the spin density
$\vec{\sigma}=\rho\,\vec{N}$ and the spin-current density
$\vec{J}_i=\rho\,v_i\,\vec{N}$ are employed.
Comparing with the quantum mechanical
results ~\cite{Li-current}, one can recognize
that $\vec{\omega}$ and $\vec{\Omega}_i$
correspond to the SU(2) gauge potentials
$\eta\vec{\mathcal A}_0$ and
$-\eta\vec{\mathcal A}_i$ with $\eta=\hbar$, respectively.
For a two-dimensional electron gas in narrow gap
zinc-blende III-V semiconductors, the SU(2) gauge potentials
have been shown~\cite{Li-current} to be related to
the Rashba~\cite{Rashba} and Dresselhaus~\cite{Dresselhaus}
SOCs, concretely,
\begin{eqnarray}\label{eq:potentials}
 \vec{\mathcal A}_x \!=\!\frac{2m}{\eta^2}(\beta,\,\alpha,\,0),
                                                    \hspace{3mm}
 \vec{\mathcal A}_y \!=\!-\frac{2m}{\eta^2}(\alpha,\, \beta,\,0),
                                                     \hspace{3mm}
 \vec{\mathcal A}_0 \!=\!0.   \hspace{5mm}
\end{eqnarray}
In terms of these gauge potentials, the SU(2) ``electric''
and ``magnetic'' fields can be expressed as
\begin{eqnarray}
\vec{\mathcal E}_i &=& -\partial_0 \vec{\mathcal A}_i
 -\partial_i \vec{\mathcal A}_0
 +\eta\vec{\mathcal A}_0\times\vec{\mathcal A}_i,
                                \nonumber\\
\vec{\mathcal B}_i &=& \epsilon_{ijk}\partial_j \vec{\mathcal A}_k
         +\frac{\eta}{2}\epsilon_{ijk}\vec{\mathcal A}_j
           \times\vec{\mathcal A}_k,
\end{eqnarray}
which provides a ``spin-related'' force~\cite{Li-current}
\begin{eqnarray}\label{eq:spinForce}
\mathcal F_i = \vec{\mathcal E}_i\cdot \vec \sigma
 +\epsilon_{ijk}\vec J_j\cdot\vec{\mathcal B}_k .
\end{eqnarray}
In the above, we have employed the notations
$\partial_0 \equiv \displaystyle\deriv{t}$ and
$\partial_i = \displaystyle\deriv{x_i}$ for simplicity.

Equations (\ref{eq:omegafields}) and (\ref{eq:continuty-like})
are the main relations of this section which depicts
a classical picture for the motion of electrons in
both U(1) and SU(2) fields.
In the following, we will carry on the discussions
on the spin transport
in which these results will be employed.

\section{Precession of the spin orientation}\label{sec:spin precession}

As an immediate application of Eq.~(\ref{eq:omegafields}),
we investigate the precession of the spin orientation
$\vec N$ in a one-dimensional quantum wire.
For example, a spin-polarized current is injected at $x=0$
and ballistically transported through the wire.
We consider the case that the  Dresselhaus SOC is homogeneous
while the Rashba SOC can be either homogeneous or inhomogeneous.

\paragraph{We first consider homogeneous $\alpha$,} the precession
of $\vec N$ can be solved analytically
and its three components are given by
\begin{eqnarray}\label{eq:const}
N^x &=& -\frac{\gamma N^z_0}{\sqrt{1+\gamma^2}\cos\theta }
          \sin(\kappa x+\theta)
          +\frac{ N^x_0+\gamma N^y_0}{1+\gamma^2},
                           \nonumber \\
N^y &=& \frac{N^z_0 }{\sqrt{1+\gamma^2}\cos\theta }
           \sin(\kappa x+\theta)
          +\frac{\gamma (N^x_0 +\gamma N^y_0) }
           {1+\gamma^2},
                           \nonumber \\
N^z &=& \frac{N^z_0 }{ \cos\theta } \cos(\kappa x+\theta),
\end{eqnarray}
where $\vec N_0=(N^x_0, N^y_0, N^z_0)$ is
the initial spin orientation,
$\gamma =\alpha/\beta$,
$\kappa = 2m\sqrt{\alpha^2+\beta^2}/\eta$
and $\tan\theta= (N^y_0-\gamma N^x_0)/
\sqrt{1+\gamma^2}N^z_0$.
\begin{figure}[ht]
\includegraphics[width=80mm]{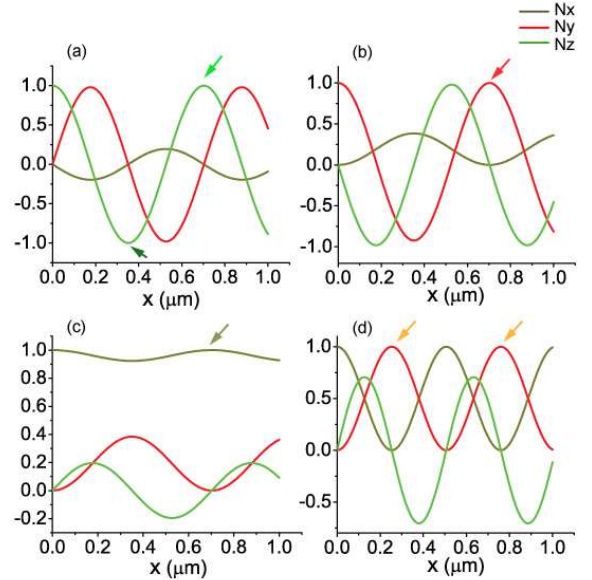}
\caption{\label{fig:const} (Color on line)
$N^x$, $N^y$ and $N^z$ are plotted in unit of $\hbar/2$
as functions of $x$   with $\alpha=10^{-12}~\textrm{eV~m}$,
$\beta=5\times10^{-12}~\textrm{eV~m}$ in panels
(a), (b) and (c). The initial direction $\vec N_0$ is along
$z$ direction in panel (a) and $\vec N$ can be antiparallel
(parallel) to $\vec N_0$ as marked by the dark (light)
green arrow. In panels (b) and (c), the initial direction
$\vec N_0$ is along $y$-axis  and $x$-axis directions, respectively.
$\vec N$ can only return its initial direction.
In panel (d), $\alpha=\beta=10^{-12}~\textrm{eV~m}$
and $\vec N$ can rotate to $y$-axis  direction at
$x$=0.25~{$\mu$m}, 0.76~{$\mu$m}
(marked by orange arrows) with
$\vec N_0$ points to $x$-axis direction. }
\end{figure}
Figure \ref{fig:const} shows the variation of
$\vec N$ with respect to $x$-axis with
$\alpha=10^{-12}~\textrm{eV~m}$ and
$\beta=5\times10^{-12}~\textrm{eV~m}$.
The initial direction $\vec N_0$ affects
the precession of $\vec N$.
When $\vec N_0$ is along $z$-axis direction,
$\vec N$ can either precess to the opposite direction
of $\vec N_0$ (\eg at $x$=0.35~$\mu$m marked
by the dark green arrow in Fig.~\ref{fig:const}(a)) or
go back to its initial direction (\eg at $x$=0.7~$\mu$m
marked by the light green arrow in Fig.~\ref{fig:const}(a)).
However, when $\vec N_0$ is along the $x$-axis or $y$-axis direction,
$\vec N$ can only return to its initial direction
and never achieve the opposite direction of $\vec N_0$,
as illustrated in Fig.~\ref{fig:const}(b) and (c).
This can be understood as follows.
If the initial direction is in the plane
perpendicular to the revolution axis
$\vec{\mathcal A}_x$, $\vec N$ will precess
in this plane and definitely experience the opposite direction of $\vec N_0$.
Besides, the position where $\vec N$ rotates to
its initial direction is determined exclusively by $\kappa$.
It can be seen from Eq.~(\ref{eq:const}) that
the three components of $\vec N$ oscillate
with the same frequency $\kappa$.
A special situation is $\alpha=\beta$
where $N^x$ and $N^y$ are formally equivalent.
Thus $\vec N$ can also precess to the $y$-axis  direction
($x$-axis direction) when $\vec N_0$ is in the $x$-axis direction
($y$-axis  direction), as shown in Fig.~\ref{fig:const}(d).

\paragraph{We consider an inhomogeneous case $\alpha=ax$}
which can be realized by tuning the applied gate voltage
on the two-dimensional electron gas. Hence,
$\vec N$ can be solved analytically in form of series expansion,
namely,
$N^x=N^x_0-\kappa_a\displaystyle\sum^{\infty}_{n=0}
\frac{b_n}{n+2} x^{n+2}$,
$N^y=N^y_0+\kappa_b\displaystyle\sum^{\infty}_{n=0}
\frac{b_n}{n+1} x^{n+1}$,
$N^z=\displaystyle\sum^{\infty}_{n=0}b_n x^n$
with $\kappa_a=2ma/\eta$
and $\kappa_b=2m\beta/\eta$.
The coefficients are given by $b_0 = N^z_0$,
$b_1=-\kappa_b N^y_0$,
$b_2 =\displaystyle\half (\kappa_a N^x_0
-\kappa_b^2 N^z_0)$,
$b_3 = \kappa_b^3 N^y_0/6$
and $b_{n+2}=-\displaystyle\frac{\kappa_b^2 b_n}{ (n+2)(n+1) }
-\frac{\kappa_a^2 b_{n-2}}{ (n+2)n }$ for $n=2,3\cdots$.
When $\alpha$ is much smaller than $\beta$,
the components of $\vec N$ can be
approximatively written as
\begin{eqnarray}
N^x &=& N^x_0+N^y_0\frac{\kappa_a}{\kappa_b \cos\theta'}
                         \nonumber \\
&& \times \Bigl[\frac{\sin(\kappa_b x +\theta')
  -\sin\theta'}{\kappa_b}-x\cos(\kappa_b x + \theta')
           \Bigr],
     \nonumber \\
N^y &=& \frac{N^y_0}{\cos\theta'}\cos(\kappa_b x+\theta'),
                         \nonumber \\
N^z &=& -\frac{N^y_0}{\cos\theta'}\sin(\kappa_b x+\theta'),
\end{eqnarray}
with $\tan\theta'=-N^z_0/ N^y_0$.
\begin{figure}[h]
\includegraphics[width=80mm]{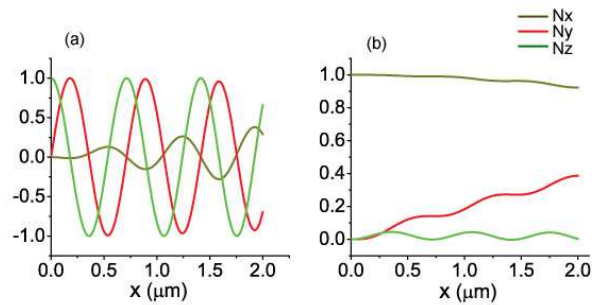}
\caption{\label{fig:linear} (Color on line)
$N^x$, $N^y$ and $N^z$ are plotted in unit of $\hbar/2$
as functions of $x$  with $\beta=5\times10^{-12}~\textrm{eV m}$
and $a=10^{-6}~\textrm{eV}$.
The initial direction $\vec N_0$ is along
$z$ direction in panel (a) and along $x$-axis direction
in panel (b). }
\end{figure}
We plot these three components as functions
of $x$ in Fig.~\ref{fig:linear} with
$\beta=5\times10^{-12}~\textrm{eV m}$
and $a=10^{-6}~\textrm{eV}$.
If the initial direction is in the $y$-$z$ plane,
$\vec N$ will almost precess in this plane
for small $x$ since the vector potential
in the $x$-axis direction $\mathcal A^x_x$
is much larger than that
in the $y$-axis  direction $\mathcal A^y_x$.
As $x$ increases, the amplitude of the oscillation
of $N^x$ becomes larger and $\vec N$ tilts
out of the $y$-$z$ plane (see Fig.~\ref{fig:linear}(a)).
If the initial direction is along the $x$-axis direction,
$\vec N$ nearly keeps pointing in this direction
for small $x$ and begins to precess as $x$ increases
(see Fig.~\ref{fig:linear}(b)).

\begin{figure}[h]
\includegraphics[width=80mm]{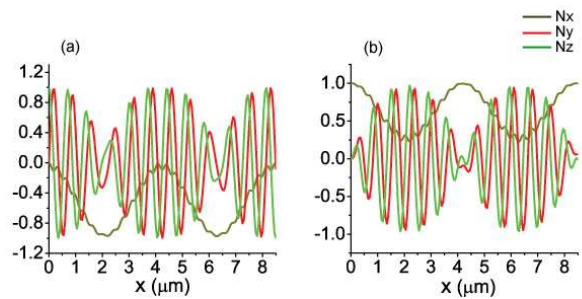}
\caption{\label{fig:sine} (Color on line)
$N^x$, $N^y$ and $N^z$ are plotted in unit of $\hbar/2$
for $\alpha=b\sin(q x)$ with $q= 10^{7}~\textrm{m}^{-1}$.
The initial direction $\vec N_0$ is along
$z$-axis direction in panel (a) and along $x$-axis direction
in panel (b). }
\end{figure}

\paragraph{We further consider sinusoid-type inhomogeneity,}
$\alpha=b\sin(q x)$.
One can solve $\vec N$ analytically
with the help of series expansion in principle,
but insufficient information can be obtained because
the solution cannot be expressed either in a closed form or
in terms of any known functions.
Now we calculate it numerically.
The obtained results show that $\vec N$
behaves like a  bi-periodic function of $x$,
one period is determined by the gauge potential
$\vec {\mathcal A}_x$ and the other by $q$,
as shown in Fig.~\ref{fig:sine}
with $b=10^{-12}~\textrm{eV m}$
and $q=10^{7}~\textrm{m}^{-1}$.

\section{Equation of motion for a single spin and its application}\label{sec:single spin}

In this section, we focus on the motion of
a single charged top $\vec{n}=(n^x, n^y, n^z)$
with constant module $n$ which is the classical analogy
of an electron. The equations of motion for this top
under both U(1) and SU(2) fields are obtained as
\begin{eqnarray}\label{eq:equationofmotion}
\frac{d\vec{n}(t)}{dt} =\eta(\vec{\mathcal A}_0
     -v^{}_i\vec{\mathcal A}_i)\times \vec{n}(t),
              \hspace{34mm}    \nonumber\\
m\frac{d v^{}_i}{dt} = \vec{\mathcal E}_i
\cdot\vec{n}(t)+e E_i +\epsilon^{}_{ijk}v^{}_j
    (\,\vec{\mathcal B}_k\cdot\vec{n}(t)+e B_k).
              \hspace{5mm}
\end{eqnarray}
The first equation can be derived from
Eq.~(\ref{eq:omegafields}) by adopting
$\vec{N}(\mathbf{r}, t) =
\vec{n}(t)\delta(\mathbf{r}-\tilde{\mathbf r}(t))$
with $\tilde{\mathbf r}(t)$ being
the trajectory of a single top.
The second equation is due to the fact
that its translational motion is governed
by both the Lorentz force and the ``spin-related'' force given
in Eq.~(\ref{eq:spinForce}).
Clearly, equation~(\ref{eq:equationofmotion}) gives rise to
some immediate consequences:
\paragraph*{Case 1}
The first equation for
the time rate of $\vec{n}$ clearly
manifests that the vector $\vec n$
does not precess when it is parallel to
$\vec{\mathcal A}_0 - v_i\vec{\mathcal A}_i $,
or more specially,
$\vec{\mathcal A}_0 - v_i\vec{\mathcal A}_i=0$.
In such cases, the spin orientation keeps unchanged,
which results in an infinite spin relaxation time.
A typical example is that the infinite
spin relaxation time occurs in the $\pm[1,\pm1,0]$
direction when $\vec{\mathcal A}_0=0$ and
$\alpha=\pm \beta$, as discussed in Ref.~\cite{SCZhang06}.
This is analogous to the case in the classical
electrodynamics where an electron moving
in the uniform orthogonal electromagnetic
fields with certain velocity,
$\mathbf{v} = \mathbf{E}\times\mathbf{B}/B^2 $,
does not feel the Lorentz force.
\paragraph*{Case 2}
In virtue of the coupling
between $\vec n$ and the SU(2) field in
the second equation of (\ref{eq:equationofmotion}),
the time rate of $\vec n$ leads to the time-dependent
effective fields even when the SU(2) fields are
time-independent.

As a concrete example, we consider
a spin Hall system with constant Rashba and Dresselhaus SOCs.
Thus $\vec{\mathcal A}_0=0$,
$\vec{\mathcal E}_i=0$,
$\vec{\mathcal B}_i=
 \frac{\eta}{2}\epsilon_{ijk}\vec{\mathcal A}_j
 \times\vec{\mathcal A}_k$
and we first neglect the electric field.
The second equation of Eq.~(\ref{eq:equationofmotion})
reduces to
\begin{equation}
m\frac{dv_i}{dt} = \eta\vec{\mathcal A_i}
 \cdot(v_j\vec{\mathcal A}_j)\times \vec{n},
\end{equation}
which gives rise to a relation between
$\vec n$ and $v_i$
\begin{equation}\label{eq:relation between n and v}
v_i(t)=-\frac{1}{m}\vec{\mathcal A}_i\cdot\vec n(t)+C_i,
\end{equation}
where
$C_i=v_{0 i}+\frac{1}{m}\vec{\mathcal A}_i\cdot\vec n_0$
are determined by the initial values $v_i(0)=v_{0 i}$ and
$\vec n(0)=(n^x_{0},n^y_{0},n^z_{0})$.
Consequently, one only needs to solve one equation,
\begin{equation}\label{eq:motion for n}
 \frac{d\vec n}{dt} =
 \eta (\frac{1}{m} \vec{\mathcal A}_i\cdot\vec n-C_i)
 (\vec{\mathcal A}_i\times\vec n),
\end{equation}
which can be explicitly written as
\begin{eqnarray}
\frac{d n^x}{dt} = -\frac{2m}{\eta} n^z
   \Bigl[\alpha C_1-\beta C_2
  -\frac{4\alpha\beta n^x
 +2(\alpha^2\!+\!\beta^2) n^y }{\eta^2} \Bigr],
  \nonumber\\
\frac{d n^y}{dt} = \frac{2m}{\eta} n^z
   \Bigl[\beta C_1 - \alpha C_2
    -\frac{4\alpha\beta n^y
   + 2(\alpha^2\!+\!\beta^2)n^x }{\eta^2}
    \Bigr], \hspace{3mm}
  \nonumber\\
\frac{d n^z}{dt} = \frac{2m}{\eta} n^x
   \Bigl[\alpha C_1 - \beta C_2
      - \frac{4\alpha\beta n^x}{\eta^2}
      \Bigr] \hspace{28mm}
       \nonumber\\
  - \frac{2m}{\eta} n^y
    \Bigl[C_1\beta - C_2\alpha
   -\frac{4\alpha\beta n^y}{\eta^2}
   \Bigr]. \hspace{24mm}
\end{eqnarray}
For $\alpha=\beta$ which is called ReD field in
Ref.~\cite{SCZhang06}, we can solve
these equations analytically:
\begin{eqnarray}\label{eq:n-0th order}
n^x \!\!&=&\!\!
 -\frac{n^z_0}{\sqrt{2}\cos\varphi}\sin(\omega t+\varphi)
 +\frac{1}{2}(n^x_0+n^y_0),
                            \nonumber \\
n^y \!&=&\!
 \frac{n^z_0}{\sqrt{2}\cos\varphi} \sin(\omega t+\varphi)
     +\frac{1}{2}(n^x_0+n^y_0),
                             \nonumber \\
n^z \!&=&\! \frac{n^z_0}{\cos\varphi} \cos(\omega t+\varphi),
\end{eqnarray}
where $\tan \varphi=(n^y_{0}-n^x_{0})/(\sqrt{2}\,n^z_{0})$
is determined by the initial conditions.
It is clear that the tip of $\vec n$ experiences
a cyclotron rotation with frequency
$\omega=2\sqrt{2}m\alpha (v_{0x}-v_{0y})/\eta$.
The instantaneous velocity solved from
Eq.~(\ref{eq:relation between n and v})
is just its initial value $v_x=v_{0x}$ and $v_y=v_{0y}$,
\ie the electron undergoes a motion with uniform velocity.
This is due to that the time-dependent parts of $n^x$ and $n^y$
only differ from each other by a minus sign.
Specially, when $v_{x}=v_{y}$,
the spin vector $\vec n$ does not precess
since $\omega=0$,
which recovers the result in Ref~\cite{SCZhang06}.

When an external electric field $\vec E=(E_x,E_y)$
is applied, which mimics to the usual spin Hall effect
in current literature, we obtain
$v_i(t)=-\frac{1}{m}\vec{\mathcal A}_i\cdot\vec
n(t)+C_i+\frac{e}{m}E_i t$.
The equation of motion for $\vec n$ is
almost the same as Eq.~(\ref{eq:motion for n})
if  $C_i$ is replaced by $\tilde C_i(t)=C_i+\frac{e}{m}E_i t$.
For the electric field is sufficiently weak,
the perturbation theory is applicable and
we can expand $\vec n$ in power series of the electric field,
$\vec n=\vec n^{(0)}+\vec n^{(1)}+\cdots$.
The $0$th-order results take the form of
Eq.~(\ref{eq:n-0th order}),
while the first order corrections are
\begin{eqnarray}
n^{x(1)} \!\!&=&\!\! -\lambda(E)~ t^2\cos(\omega t+\varphi),
                                         \nonumber \\
n^{y(1)} \!&=& \lambda(E)~ t^2\cos(\omega t+\varphi),
                                         \nonumber \\
n^{z(1)} &=& -\sqrt{2} \lambda(E)~ t^2\sin(\omega t+\varphi),
\end{eqnarray}
with $\lambda(E)=e\alpha n^z_0(E_x-E_y) /(\eta\cos\varphi)$.
The form of the entire motion for $\vec n$
is similar to the case without external electric fields,
but with a time-dependent amplitude
$\tilde a(t)=[(n^z_0)^2/(2\cos^2\!\varphi)
+\lambda^2 t^4]^{1/2}$ and a phase
$\tilde \varphi(t)=\varphi+\tan^{-1}(\sqrt{2}
\lambda t^2\cos\varphi/n^z_0)$.

\section{Diffusion equations and Zitterbewegung effects}
\label{sec:diffusion equation}

In the presence of U(1) and SU(2) fields
$ mv_i=p_i(\mathbf r, t)-eA_i -\eta\vec{\mathcal A}_i\cdot \vec N$,
we can write out the charge current $e\rho v_i$
and spin current $\rho v_i\vec N$ as follows
\begin{eqnarray}\label{eq:currents}
 ej_i = \frac{e\rho}{m}(p_i-eA_i) -\frac{e\eta}{m}
   \vec{\mathcal A}_i \cdot \vec\sigma,
           \nonumber \\
 \vec J_i= \frac{\vec\sigma}{m}(p_i-eA_i)
  -\frac{\eta}{m} \vec N
  (\vec{\mathcal A}_i \cdot \vec \sigma).
\end{eqnarray}
The first terms on the right hand sides of
Eq.~(\ref{eq:currents}) corresponds
to the conventional non-relativistic currents.
It is worthwhile to pay attention to the second terms
which are related to the \emph{Zitterbewegung}.
It was ever believed that the
\emph{Zitterbewegung}~\cite{Schrodinger}
as a relativistic effect cannot be observed directly
due to the high frequency (of order $10^{20}$ Hz) and short length scale
(of order $1$pm)~\cite{Huang}.
Recently, J. Schliemann \etal suggested
to detect the \emph{Zitterbewegung} in III-V semiconductor
quantum wells~\cite{Loss} where the energy and length scale
become available in experiments.
We will show that the existence of
\emph{Zitterbewegung}-related phenomena in the spin transport
makes the diffusion equations for the spin
and charge densities couple to each other.

In comparison to the expression of
the conventional non-relativistic current
density, there is an extra term
$j'_i=[\frac{e}{m}\nabla\times\vec\sigma ]_i $
in the expression derived from the Dirac equation
in the non-relativistic limit.
This term $j'_i$ can be regarded  as a result of the
\emph{Zitterbewegung}~\cite{Huang,Recami}.
As long as we distinguish between the spin space and
the conventional spatial space, we have
\begin{eqnarray}\label{eq:Zitter}
j'_i=\frac{e\eta}{m}\vec{\mathcal A}_i \cdot\vec\sigma
    + \frac{e}{m}[\nabla\rho\times\vec N]_i.
\end{eqnarray}
In the above derivation,
we have adopted Eq.~(\ref{eq:omegafields}) and
the fact that the matrix
$(\mathbb A)_{ij}\equiv \mathcal A^j_i$ is traceless.
Hence we can consider $\frac{e\eta}{m}\vec{\mathcal A}_i
\cdot \vec\sigma$ arising from the \emph{Zitterbewegung}
for either uniform $\rho$ or $\nabla\rho\propto\vec N$.
We will see in the following that
$\frac{e\eta}{m}\vec{\mathcal A}_i \cdot \vec\sigma$
enters into the diffusion equations for
both charge and spin densities
which couple to each other.

The coupled charge-spin diffusion equations
in a spin Hall system have been investigated
by several groups in the view of quantum
kinetic theory~\cite{diffusion}.
Starting from the Boltzmann equation as well as
the above results, we can derive the
semiclassical diffusion equations for
both charge and spin densities.
The introduced distribution function $f(\mathbf r, \mathbf p, t)$
obeys the Boltzmann equation
\begin{eqnarray}\label{eq:Boltzmann}
\frac{\partial f}{\partial t}+v_i\frac{\partial f}{\partial x_i}
 +F_i\frac{\partial f}{\partial p_i}
 = \frac{\partial f}{\partial t}\vert_{\textrm{coll}}.
\end{eqnarray}
For charged top, $F_i$ should contain both the Lorentz force
caused by the U(1) fields $E_i$ and $B_i$ as well as
the ``spin-related'' force caused by the SU(2) fields
$\vec{\mathcal E}_i$ and $\vec{\mathcal B}_i$.
The right-hand-side of Eq.~(\ref{eq:Boltzmann})
represents the collision term contributed
by the elastic scattering between electrons.
For the system not driven far away from the equilibrium,
the relaxation-time approximation is applicable
and the distribution function $f$ can be decomposed into
the equilibrium distribution $f_0$ and a small deviation $f'$,
in which $f_0$ is independent of the direction of $\mathbf p$.
For the spin Hall system with constant Rashba and
Dresselhaus couplings, the magnetic field and
the SU(2) ``electric'' field vanish and
$\epsilon_{ijk}v_j\vec{\mathcal B}_k \cdot \vec N
\frac{\partial f'}{\partial p_i}$ is of the 2nd order
and can be neglected.
Hence equation~(\ref{eq:Boltzmann}) is written as
\begin{eqnarray}\label{eq:charge-diff}
 \frac{\partial f}{\partial t}
 +\widetilde\nabla_i (\frac{p_i}{m}f)
 -\frac{\eta}{m}\vec{\mathcal A_i}
 \cdot\frac{\partial \vec g}{\partial x_i}
 = -\frac{f-f_0}{\tau},
\end{eqnarray}
where
$\widetilde\nabla_i =\displaystyle \nabla_i
 + e E_i \frac{\partial}{\partial\varepsilon} $
with $\varepsilon=\displaystyle\frac{p^2}{2m}$
and $\vec g=f\vec N$ is the distribution
function for the spin density.

Integrating Eq.~(\ref{eq:charge-diff}) over
$\mathbf p$, we have the diffusion equations for the charge
density $\rho=e\int f d\mathbf p$.
In the calculation, we encounter
$\int (\frac{p_i}{m}f)d\mathbf p =\int(\frac{p_i}{m}f')d\mathbf p$.
For a steady state, $f'$ can be solved from
Eq.~(\ref{eq:charge-diff})
\begin{eqnarray}
f'=-\tau(\frac{p_j}{m}\widetilde\nabla_j f
         -\frac{\eta}{m}\vec{\mathcal A}_j \cdot \vec N
          \frac{\partial f}{\partial x_j}).
\end{eqnarray}
Neglecting higher orders of $f'$, we have
\begin{eqnarray}
 \int d\mathbf p~(\frac{p_i}{m}f')
 \simeq -\tau \delta_{ij}\frac{\varepsilon}{m}
 \widetilde\nabla_j \int d\mathbf p f_0
 = -D\widetilde\nabla_i \rho.
\end{eqnarray}
Here $D=\varepsilon \tau$ is the diffusion coefficient
which turns to be  $ v^2_F \tau/2m$ in the quantum kinetic equation with
$v_F$ being the Fermi velocity.
As a result, we have
\begin{eqnarray}\label{eq:couple1}
 \der{\rho}{t} -D \widetilde\nabla^2 \rho
 -\frac{\eta}{m} \vec{\mathcal A}_i
   \cdot\frac{\partial \vec \sigma}{\partial x_i}
 =-\frac{\rho-\rho_0}{\tau},
\end{eqnarray}
and the charge current is given by
\begin{eqnarray}
J_i=\int d\mathbf p~v_i f'
   =-D\widetilde\nabla_i \rho
    -\frac{\eta}{m} \vec{\mathcal A}_i\cdot\vec \sigma.
\end{eqnarray}
We derive the diffusion equation for the spin density
in an analogous procedure,
\begin{eqnarray}\label{eq:couple2}
 (\frac{\partial}{\partial t}
 -\eta\vec{\mathcal A}_0\times)\vec\sigma
 -D (\widetilde\nabla_i
 +\eta\vec{\mathcal A}_i\times)
 (\widetilde\nabla_i
 +\eta\vec{\mathcal A}_i\times)\vec\sigma
                       \nonumber \\
  -\frac{\eta}{m} (\vec{\mathcal A}_i\cdot\vec N)
    \vec N \frac{\partial \rho}{\partial x_i}
 =-\frac{\vec\sigma-\vec\sigma_0}{\tau},
\end{eqnarray}
with the spin current being
\begin{eqnarray}
\vec J_i= -D (\widetilde\nabla_i
  +\eta\vec{\mathcal A}_i\times)\vec\sigma
  -\frac{\eta}{m} (\vec{\mathcal A}_i\cdot\vec N)
    \vec \sigma.
\end{eqnarray}
Eq.~(\ref{eq:couple1}) and Eq.~(\ref{eq:couple2})
are classical counterparts of the coupled
spin-charge diffusion equations~\cite{diffusion}.
They are valid for a large time scale
since our startpoint is the semiclassical
Boltzmann equation which implies the assumption of
the energy conservation.
For a short time scale $\delta t$,
$\delta \varepsilon$ is large and
the quantum kinetic theorem should be employed.

The last terms on the left hand sides of
both Eq.~(\ref{eq:couple1}) and
Eq.~(\ref{eq:couple2}) are the
contributions from the \emph{Zitterbewegung}
which make these diffusion equations
coupled to each other.
To illustrate the effect of the
\emph{Zitterbewegung}-related contributions,
we consider the one-dimensional quantum wire
with constant Rashba and Dresselhaus SOC.
The charge density exponentially decays
in the wire, namely, $\rho(x)\propto e^{-x/Ls}$
where the decay length is given by
$L_s=(\sqrt{\xi^2+4D/\tau}+\xi/(2D))^{-1}$ with
$\xi=2(\alpha^2-\beta^2)(\alpha N^y_0+\beta N^x_0)
/\eta(\alpha^2+\beta^2)$,
shorter than that without the
\emph{Zitterbewegung}-related contributions
$L_s=(\sqrt{D\tau})^{-1}$.

\section{Summary}\label{sec:summary}

Depicting moving electrons as a continuum constituted by charged tops,
we investigated the motion of electrons in both U(1) and SU(2)
fields and discussed the spin transport.
We derived the classical counterpart of
the continuity-like equations for the spin current,
which takes the same form as that ever proposed
quantum-mechanically in a previous paper~\cite{Li-current}.
This provided us an intuitive picture for elucidating
the non-conservation of the spin current in the presence of the SOC.
We discussed the precession of the spin orientation in a ballistic
one-dimensional quantum wire with both Rashba and Dresselhaus SOCs
and found that the initial direction can greatly affect the spin precession.
We also formulated the equations of motion for a single spin
in the presence of both U(1) and SU(2) fields.
As a direct consequence, we obtained the condition
for the emergence of infinite spin relaxation time.
A special situation of our conclusion recovers
the previous result discussed by other authors.
Furthermore, we derived the semiclassical
diffusion equations for both the charge and spin densities.
We found that the \emph{Zitterbewegung} causes
these equations coupled each other and
 makes the decay length of the charge density much shorter
in a one-dimensional quantum wire.

\acknowledgments The work was supported by NSFC Grant No. 10674117
as well as  by PCSIRT Grant No. IRT0754.


\begin{references}
\bibitem{footnote}
A complete  quantum mechanical description is given by four local quantities
$\psi^*_\uparrow \psi_\uparrow$, $\psi^*_\uparrow \psi_\downarrow$,
$\psi^*_\downarrow \psi_\uparrow$, and $\psi^*_\downarrow \psi_\downarrow$,
of which the proper compositions define one probability
density and three components of  spin densities,
whose classical counterparts are $\rho$ and $\vec{N}$.
\end{references}
\end{document}